\documentclass[twocolumn,showpacs,showkeys%
,amssymb,amsmath,nobibnotes,aps,prd]{revtex4}
\usepackage{amsmath}%
\usepackage{graphicx}%
\begin{document}
\title{Theory of superconductivity of gravitation and the dark matter enigma}
\author{Wenceslao \surname{Santiago-Germ\'{a}n}}
\affiliation{Manuel Sandoval Vallarta Institute for Theoretical Physics,  Oth\'on P. Blanco 40, C.P. 77098, Quintana Roo, M\'exico }
\email{wsantiago@uqroo.mx}
 \begin{abstract}
In this article, the question of the nature of cold dark matter is approached from a new angle. By invoking the Cauchy problem of relativity it is shown how---under very precise astrophysical conditions---the Einstein general theory of relativity is formally equivalent to the Ginzburg-Landau theory of superconductivity. This fact lead us to suspect that the superconductivity of gravitation ought to be a real physical process occurring in the outskirts of galaxies. It is found that quantum mechanically gravity can achieve a type-II superconductor state characterized by the Gizburg-Landau parameter $\kappa=1.5$, and it is suggested that a probability flux of Cooper pairs (quantum gravitational geons charged with vacuum energy) are directly responsible for the flatness exhibited by the rotation curves in spiral galaxies, as well as the exotic behaviour observed in galactic cluster collisions. If this hypothesis proves correct, the whole phenomenon of dark matter may count, after all, as another triumph for Einstein's theory of gravity. The tension between gravitation and quantum mechanics is explored further by a subtle consideration of  the Hamilton-Jacobi theory of the York-time action---providing additional motivation for the above line of reasoning. In particular, Penrose's estimate for the rate of collapse of the wavefunction is recovered, and connected to the instability of Misner space.  
\end{abstract}
\maketitle
  \section{\label{sec:intro}Introduction.}
  In cosmology---in order to not overthrown Newtonian mechanics in the non-relativistic regime (embracing large distances and small accelerations\cite{PeeblesCosmology})---an invisible substance has been postulated: the mysterious cold dark matter. According to current theories it holds the key to unravel the inner workings of the formation and stability of the large-scale structure of the universe\cite{Rees, Blumenthal}. Furthermore, approximately $22.7\pm1.4 \%$ of the cosmos total mass-energy density should be in the form of cold dark matter\cite{Jarosik}.  Consistency with astrophysical data requires cold dark matter to be made of  electrically neutral, QCD colourless, massive particles: in a cold\cite{Blumenthal, Peebles1983}, stable (or long-lived), unexcited state. In particular, these hypothetical particles are pictured orbiting the outskirts of luminous galaxies\cite{Rubin} and flowing without resistance in galactic cluster collisions---like the one detected in the double galaxy cluster 1E 0657-558: the `bullet cluster'; thus,  they must have a negligible non-gravitational  interaction with ordinary baryonic matter or themselves\cite{Clowe}. However, nobody knows for sure their  exact nature or even if they exists at all, since it has been stated that a modification to Newtonian dynamics or gravity may account for the same effects\cite{Angus, BM, TeVeS}. The purpose of this article is to show that the highly non-linear Einstein law of gravity is---under very precise astrophysical circumstances---formally equivalent to the Ginzburg-Landau theory of superconductivity\cite{G-L, Abrikosov},   throwing  light on the dark matter enigma. 
  
 In particular, it is found that quantum mechanically gravity can achieve a type-II superconductor state characterized  by the constant $\kappa=1.5$. For many years it has been speculated\cite{WeinbergCosmology} that the axion\cite{Peccei, Preskill} or the lightest supersymmetric particle\cite{Cabibbo, Ellis} (still undetected\cite{Supersymmetry}), or perhaps a radical departure to: the law of gravity\cite{Bekenstein, Moffat}  or Newtonian dynamics\cite{Milgrom}, might solve the cold-dark-matter puzzle. In contrast, it is found here that the four-dimensional Einstein's field equations themselves suggest that nonbaryonic  cold dark matter might consist of a probability flux of Cooper pairs\cite{BCS}: quantum gravitational geons\cite{Wheeler1957} charged with vacuum energy, orbiting the outskirts of luminous  galaxies and modifying the gravitomagnetic lines of force. If this hypothesis proves correct, and if dark energy turns out to be only a manifestation of Einstein's cosmological constant, the whole phenomenon of dark matter may count, after all, as another triumph for Einstein's theory of gravity. 

Historically, Bryce DeWitt was the first to point out, in 1966, a direct connection between gravitation and superconductivity (in metals),  when he showed by calculation that the magnetic field inside a mettalic superconductor is non-vanishing whenever a Lense-Thirring field is present, and that it is the flux of a linear combinations of the magnetic and Lense-Thirring fields which gets quantized in virtue of the Copper pairs\cite{DeWittSuper}. Much later, a close analogy between the superconductors of the second kind and the Einstein theory of gravity was noticed by P. O. Mazur, who argued that the spacetime around of a `spinning cosmic string' can be regarded as a gravitational analog of the Aharonov-Bohm solenoid or Abrikosov vortex\cite{Mazur}.  In this analogy, the angular momentum per unit length corresponds to the magnetic flux, and the mass-energy per unit length corresponds to the charge. Near to the axis of the string, however, closed timelike curves appear, exactly at the radius where the velocity of frame dragging exceeds the speed of light. Nevertheless a quantization rule on the energy of particles propagating on this background  was found for which the spinning string cannot be detected by scattering experiments. This work lead to a nonrelativistic superfluid condensate model for an emergent spacetime\cite{Chapline}. More recently, (holographic) guage/gravity duality arguments have been applied to obtain a gravitational dual description of some aspects of type II superconductivity in 2+1 non-gravitational systems\cite{Hartnoll}, including vortex configurations\cite{Johnson} and the Josephson effect\cite{Horowitz}. The bulk is taken to be a four-dimensional, electrically charged, AdS black hole  with planar horizon geometry, that develops non-trivial scalar hair at low temperatures. Such physical configuration behaves like a thermal system in one lower dimension, where  the role of charged condensate is played by the charged scalar field, and the temperature of the superconductor is given by the Hawking temperature of the black hole. 

 In the present article the notion of superconductivity---experimentally discovered at $4.12^\circ$ by  Kammerlingh Onnes nearly a century ago---is extended into the realm of gravitation: not in the form of an analogy, but as a basic feature of the quantum physical properties of a dynamical four-dimensional spacetime, resulting in new effects with potentially testable consequences, such as the formation of quantum vortices and the physics of the gravitational Meissner effect.
 
The plan of the paper is as follows. In  Sec. \ref{CDMM} a formulation of the cold dark matter puzzle is presented, which promotes revisiting the question of the origin of inertia from a quantum mechanical perspective. That the superconducting state of gravitation might be a real astrophysical process is hinted in Sec. \ref{super}. To support the case, in Sec. \ref{GLsection},  it is shown how under precise circumstances the Cauchy problem of general relativity is connected with type-II superconductivity. A variational approach regarding this central issue is developed in  Sec.\ref{Vsection}. Further insight is gained in Sec. \ref{HowCome}, where Wheeler's old conundrum about the origin of the quantum is considered, albeit in a modest way. This is done by developing mathematical relationships intrinsic to the general theory of relativity, which however are subjected to a quantum-mechanical interpretation. Indeed, as it will be shown, the proposed mathematical scheme drives one ineludibly to  a crossroad  where the two other intricacies of contemporary physics; namely, space-time singularities and the measurement paradox of quantum mechanics, converge or meet in a subtle way.  The implications of the theory, in its present primitive stage of development, are considered in Sec. \ref{ConclusionSection}, where the basic results are discussed and summarised.

\section{The mystery of cold dark matter: The quanta of Mass-energy `there' rules inertia `here'}\label{CDMM}  
According to the prevailing view  \cite{Einasto1974},  at extragalactic scales the expanding universe is best think of as consisting of two parts: One luminous---obeying Newtonian mechanics in the limit of slowly moving bodies and large distances, and the other dark---which is several times more abundant than the first one, and from which the formation and stability of the large scale structure of the universe rests upon. The quality of being  invisible (or dark) is bring at front since it is only through its  gravitational interaction with other bodies that this hypothetical form of matter has been  (so far) accounted for. Thus---in case it exists---it should not  have both electric charge and QCD colour, but it should  posses a local (or non-local) mass. Luminous galaxies are pictured as if they were embedded in extensive cold dark `halos' (out to $80$ kpc in some cases, approximately $5$ to $10$ times more massive than the observed  luminous mass\cite{Rees}, and whose structure (density profile) is inferred and tested with the help of numerical simulations\cite{Press}). This simply {\it hypothesis} is first and for most based on the observation of an anomalous velocity dispersion of galaxies within cluster of galaxies\cite{Zwicky}, as well as on the non-Keplerian motion  (the existence of  very extensive neighbourhoods of constant velocity flow) of hydrogen clouds outside the bright parts of spiral galaxies \cite{Rubin1986}. Curiously enough, the velocities involved in these physical processes are highly nonrelativistic\cite{Blumenthal}.  Furthermore, using  gravitational lensing and X-ray data it has been inferred  that, during the merger of two galactic clusters, galaxies behave relatively simple, in view of the fact that they act as  collisionless particles that spatially decouple from the fluidlike X-ray-emitting intracluster plasma that experiences ram pressure\cite{Clowe}. Therefore, cold dark matter does not appreciably interact---except through gravity---with ordinary baryonic matter or itself.  This can be used as hard evidence supporting the view that most of the cold dark matter in the universe is nonbaryonic,  a conclusion which is also required to not enter into conflict  with primordial big bang nucleosynthesis\cite{helio} and the observed residual lithium, deuterium, and helium-3 abundances\cite{Nucleo, Gott}. And it is precisely at this point  (as the title of a famous short story dictates: `the garden of forking paths') that one might decide to go outside the realm of well established theory: to point out the true identity (or multiple identities) of such  exotic nonbaryonic particles.  In this paper we shall try to resist such an impulse, but  particle physics---through  various extensions of the standard model---indeed offer such an opportunity by providing us with both, a seductive line of thought and a  large list of cold dark matter candidates: including the lightest supersymmetric particle predicted by R-parity-conserving supersymmetry (which if it is not the gravitino, it could be either a sneutrino or a neutralino---which are typical `WIMPs', Weakly Interacting Massive Particles) and axions (postulated to solve the strong CP puzzle) which are pseudo Goldstone-bosons. 
   
   In principle these WIMPs candidates---assuming they exist---might be produced in the laboratory or detected 
(through their elastic scattering with nuclei) from the halo of the Milky Way that pass through the laboratory (say located very deep inside a mountain)\cite{LSmith}.  So far no clear-cut evidence for a WIMP signal has been found that has been corroborated by two independent laboratories,  although there has been some improvements in the limits for the existence and detection of cosmic WIPMs by collaborations such as CDMS-II, CoGeNT, CRESST, DAMA/LIBRA, EDELWEISS-II,  HDMS, ORPHEUS, PICASSO, UK Dark matter, WARP, and XENON100---which confront the problem of distinguishing true WIMP events from background caused by natural radioactivity and cosmic rays, and from glitches in their electronics\cite{CDMS, XENON100}. In principle, WIMPs can also be detected indirectly though the observation of other particles produce when pairs of WIMPs annihilate.  In a recent finding, involving collected data from the Fermi Gamma-ray space telescope and the analysis of seven dwarf spheroidal galaxies in the vicinity of the Milky Way (Bootes I, Draco, Fornax, Sculptor, Sextans, Ursa Minor, and Segue 1), it was realized that generic WIPMs candidates  annihilating into $\bar{b}b$ with mass $m_w$ less than $40$ GeV cannot be dark matter particles\cite{Geringer},  demanding a revision of certain claims of WIMP detection by underground experiments.  

 The axion can be converted into photons by intense magnetic fields, this fact has been exploited  to impose cosmological and astrophysical limits to their mass  which it is expected to be in the range of $10^{-5}-10^{-2}$ eV \cite{Zioutas}.   Primordial black holes, steril neutrinos, little Higgs particles, axinos, and the lightest Kaluza-Klein particle figure as other viable cold-dark matter candidates. But let us suppose that the above point of view is turned up side down, say by rejecting all the way the existence of cold dark matter, then one is lead to more radical proposals tempering with the very own structure of Newtonian mechanics in the advent of small accelerations (of the order of $1.2 \times 10^{-10} m s^{-2}$) or with Newtonian gravity at large distances \cite{TeVeS, NewTeVeS}, which however give excellent fits to the rotations curves and allow a direct derivation of the Tully-Fisher relation.  We shall not, however, follow that path either. 
 
 Both, the supersymmetric particle hypothesis and the nonstandard  kinematics, offer a world view that has not yet been contradicted---or confirmed---by experiments,  so the question remains: Does cold dark matter exists at all? Keeping as needed the luminous galaxies and cluster of galaxies in bound stable states?---or it does not exist,  but then: How on earth our  theories have been misapplied? 
 
 Let us state clearly that we shall stick all the way with the basic nonlinear field equations of Einstein's theory of gravity. The point of view adopted here is that gravity has `a lot' to say about why the quantum theory is the way it is. Quantum mechanics is think of as been interconnected with---or perhaps even ruled by---gravity in a subtle way. This might come as a surprise by the easiness one runs into trouble when a direct, tour de force approach, is used to explore a possible a union between the two of them, but one should take in mind that there are some `serious' physical questions left aside by the mathematical formalism of quantum mechanics: the shifty split between micro-macro, reversible-irreversible, and quantum-classical, as was unceasingly stressed by Erwin  Schr\"{o}dinger  (and later on by John Bell) and vividly encapsulated by the cat-measurement  paradox. Does gravity offers a way out to these often ignored `ontological questions'?
  
    In our view, the mystery of cold dark matter is a symptom of a  bigger crisis than the one usually cured by just adding a new type of particle:
   
 {\it The failure of a proper  understanding of how the quanta of mass-energy  `there' rules inertia `here.' }
 
Indeed much is gained by flipping from the dark matter perspective  to the realm of quantum gravitational phenomena, since there is now---as Hilbert could have put it, ``a  guide post  on  the mazy paths of  hidden truths,"  for quantizing the gravitational field. ``Quantum gravity is a very tough problem,'' warned  W. Pauli to B. S. De Witt \cite{PauliDW}:  How are we going  to unify the strange world of Max Born's {\it probability wave amplitudes}, $\psi${\it 's},  with the peculiarities of the Einstein's four-dimensional curved space-time continuum? 
 
Perhaps we have various clues already:  

{\it There is an electrically neutral, QCD colourless, quasi-substance with local (or non-local mass)  that is in a cold, stable (or long-lived) unexcited state far away of any luminous zone and strong field; it flows freely (without resistance) but only at non relativistic speeds---as if there  were a limiting velocity that it cannot surpass, it has a negligible  non-gravitational interaction with ordinary baryonic matter or itself.}  What could it be? 

To cope with the subtleties imposed by the above scenario let us turn to mathematics since as Max Born put it \cite{MaxBorn1954}: ``when in conflict, mathematics---as often happens---is cleverer than interpretative thought.''
  
\section{From cold dark mater to superconductivity}\label{super}  
  Superconductivity was the expression used by H. K. Onnes\cite{Onnes} to describe his discovery of an abruptly lost of current resistance in metals at low temperatures, and it was  shown to be more startling than expected, as new properties: the Meissner  effect\cite{Meissner}, the quantization of flux\cite{London}, and the {\it ac} Josephson effect\cite{Josephson},  were exposed leading to more complete picture of the mechanism responsible for superconductivity. The Bardeen-Cooper-Schrieffer (BCS) theory, proposed in 1957,  provides the essential features behind the microscopic explanation of superconductivity in metals\cite{BCS}. It states that if the temperature of a metal is sufficiently low,  then conduction electrons (with opposite momenta and spins) near the Fermi surface may become bound in pairs, by an attractive force (however small) coming from the interaction with the vibrations of the lattice. The mettalic superconducting current is then pictured as being formed by the so called Cooper pairs (of zero spin; $S=0$) which become coherent, i.e. described {\it by the same} low energy wave function. Historically, the idea of pairing was hinted in the works of R. A. Ogg and M. R. Schafroth\cite{Ogg, Schafroth}.
 
    To see how superconductivity and the associated wave-particle duality might arise in pure gravity let us list five evocative facts:
    
     {\it First,} it is curious and interesting that Rubin's discovery---of an almost constant nonrelativistic velocity flow $v$ of hydrogen clouds outside the bright parts of spiral galaxies---smoothly fix the Newtonian gravitational potential  $\phi=-GM/r$ to a constant value (where $M$ is the mass within radius $r$ and $v$ is typically  (ref.\cite{Rubin}) of the order of $100-300$ km/s);  meaning that  
    \begin{equation}
  \Psi \equiv 1+(2\phi/c^2)\approx 1-2(v/c)^2\approx cte., 
    \end{equation}
   over an extended (out to $\sim$ $80$ kpc in some cases\cite{Rees}) ring-shaped region of space.  An astonishing similar constrained dynamics arises in the context of superconductivity, where  the stiffness of ``the wave function $\Psi$'' results from the appearance of an energy gap $\Delta_{\mathcal{N-S}}$ (computed below) between the energies of the first excited state and the ground state\cite{Feynman1972}. 
   
   {\it Second,} it has been known for a long time that in the gravitating field of a spherical rotating mass, the geodesic motion (in the post-Newtonian approximation) is ruled by a Lorentz-like force, where mass plays the role of electric charge\cite{WheelerCIU}: 
 \begin{equation}   
    m^{\ast}\frac{d}{dl}[(1+ \phi )\mathbf{v}]  \approx m^{\ast} (\nabla\phi +  \frac{\partial \mathbf{A}}{\partial l} )+  m^{\ast} \mathbf{v} \times (\nabla \times \mathbf{A}),
    \end{equation}
  here 
\begin{equation}  
  \phi = \frac{\chi}{8\pi}^{-1} \int \frac{\rho} {r} dV_o,
 \end{equation} 
 and 
\begin{equation} 
\mathbf{A}=\frac{\chi}{8\pi} \int  \frac{ 4\rho  \mathbf{v}} {r} dV_o.
\end{equation}
The term $m^{\ast}\nabla\phi$ gives the Newtonian force of gravity, $\rho$ is the local energy density, $\mathbf{v}$ is the velocity. In this approximation,  $\mathbf{A}$ is the gravitomagnetic potential and $\mathbf{H}=\nabla \times \mathbf{A}$ the gravitomagnetic field. A factor four affects the $ \mathbf{A}$-formula. Intuitively {\it every mass $m^{\ast}$ can be regarded as a pure imaginary charge,} so that masses of equal sign attract each other. 
\begin{eqnarray}
m^\ast \mapsto  &  ie^\ast\sqrt{1/ 4\pi \epsilon_o G}.
\end{eqnarray}
This way of thinking will prove fruitful later to account for some internal symmetries that are present in the theory when there is a positive cosmological constant. The operation of complex conjugation will be used to account for its repulsive nature. 
\begin{eqnarray}
 \sqrt{\Lambda/3} \mapsto & -i\frac{e^\dagger}{c\hbar} \sqrt{ 1/4\pi \epsilon_oG} c^2,
\end{eqnarray}

{\it Third,} it has been acknowledge that the indefiniteness of the gravitational action in the path-integral approach to quantum gravity implies that the conformal features of the metric must be handle with care\cite{Hawking1979}. 

{\it Fourth,}   sure enough, the scheme by H. Weyl to unite general relativity with electromagnetism\cite{Weyl}, when applied to the gravitomagnetic potential $\mathbf{A},$ provides a natural way to introduce complex numbers in Einstein's theory: via gauge  transformations and  conformal scalings 
  \begin{eqnarray}
      \partial_j  & \mapsto &  \partial_j-\frac{iq}{c \hbar}A_j   \label{Avector} \\
  A_j   &\mapsto &  A_j -\varphi_{,j};  \\
   g_{\mu \nu} & \mapsto &   \exp({- \frac{   i q \varphi}{ \hbar c}})    g_{\mu \nu}.  \label{scaling}  
 \end{eqnarray}
 where  $q$ is some charge and $\varphi$ an scalar. In Weyl's scheme the internal symmetries of the electromagnetic radiation field---expressing the interchangeability among the electromagnetic potentials that can occur at a single spacetime point,  are regarded as geometrical symmetries---expressing the interchangeability of points of spacetime, by an appropriated  rescaling of the metric. 
 
  {\it Five,}  in 1998  S. Perlmutter, B. P. Schmidt, and A.G. Riess  through observations of distant Type 1a supernovae  discovered that the universe is expanding at an accelerated rate\cite{Riess, Perlmutter}. The fate of the cosmos hangs; therefore, on an unknown physics: `The one' responsible for giving the cosmological constant  $\Lambda<3\times10^{-52}m^{-2}$ its actual {\it nonzero} value. 
 
 \section{Gravitation and the Gizburg-Landau theory of superconductivity}\label{GLsection}
  In 1950,  Landau and Ginzburg introduced their semiphenomenological theory for superconductivity\cite{G-L}, which  is based on the general theory of second order phase transitions of Landau\cite{PhaseT}, developed around 1937. In this theory a sort of macroscopic  wave function ``$\Psi$''  is used as an order parameter (which is finite below the transition and zero above it).  By 1959,  L. P. Gor'kov\cite{Gor'kov}  showed how the (GL) equations for superconductivity in metals   can be derived from the (BCS) equations in the case of a short range potential near the critical temperature of the superconductor (a rigorous and more recent mathematical treatment can be found in ref.\cite{Frank}). The complete set of  Gizburg-Landau equations of the theory of superconductivity is given by the following relations. First, the  (GL) equation
 \begin{equation}
 \frac{1}{2 m^\ast}(\frac{\hbar}{i}\nabla-\frac{e^{\ast}}{c}  \mathbf{A})^2 \Psi +\alpha\Psi + \beta |\Psi|^2 \Psi=0, \label{GLequation}
  \end{equation}
 where it was found experimentally, in the case of superconductivity in metals, that $e^\ast$ is twice the electric charge of the electron; the coefficients $\alpha$ and $\beta$ are material parameters which are deduced from subsequent measurements and may depend on temperature. 
  Second, the Maxwell's equation  
\begin{equation}
\bigtriangleup \mathbf{A} = -(4\pi /c) \mathbf{j}_s, \label{Awave}
\end{equation}
which is satisfied when in a stationary situation  the vector potential $\mathbf{A}$ obeys the Lorentz gauge; and finally 
 an expression for  $\mathbf{j}_s$ which reduces to  $e^{\ast}$ times the probability current density: 
\begin{equation}
\mathbf{j}_s=\frac{\hbar e^{\ast} }{i 2 m^\ast }(\Psi^{\ast}\nabla \Psi- \Psi \nabla \Psi^{\ast})- \frac{{e^{\ast}}^2}{m^\ast c } |\Psi |^{2} \mathbf{A}.
\end{equation}
 To encompass the various thoughts already expressed, about the nature of cold dark matter, into a single mathematical expression, one that is intrinsic to the initial-value problem of general relativity (the appropriated setting for studying the  origin of inertia \cite{WheelerCIU}), consider the  generalisation of the  Newtonian scalar potential $\Psi$ satisfying the highly non-linear Lichnerowicz equation\cite{Lichnerowicz}. $\Psi$ is set to conformally  deform  the `physical' 3-metric $g^{ij}$ to a more primitive one $\tilde g^{ij}, $ taking 
 \begin{equation} 
   g^{ij}= \Psi^{-4} \tilde g^{ij}.
 \end{equation}
  By construction $\Psi$ {\it is  positive and no where zero, so it might be regarded as a probability wave function describing a state of lowest energy:}  
  
  The Lichnerowicz equation, introduced to gravitation in 1944, is given by 
   \begin{eqnarray} 
  (&-&8 \tilde\bigtriangleup - \tilde M|\Psi|^{-8}) \Psi + \,^{(3)}\tilde R \Psi  \nonumber \\
 &+&  (2/3)(T^2|\Psi|^2) |\Psi|^2\Psi - \tilde Q |\Psi|^{-4}\Psi  =0, \label{Lequation}
  \end{eqnarray} 
and it is nothing more than the Hamiltonian constrain of general relativity written in a clever way\cite{WheelerCIU}.  $\tilde M,$   $\,^{(3)}\tilde R,$ $T,$  and $\tilde Q,$ are respectively the {\it conformal} density of  gravitational-wave effective kinetic energy, the Riemann scalar curvature invariant of the {\it conformal} 3-space metric,  the York time $T$ (geometrically the trace of the extrinsic curvature: $T=$ Tr $K$ on a spacelike hyersurface), and  the {\it conformal} local energy density of ordinary mass-energy $\tilde Q= 2 \chi  \tilde\rho.$ 
   The momentum constrain of Einstein's gravity theory is given by 
    \begin{equation}
 \tilde\bigtriangleup^{\star} W_i =  8\pi \tilde{j}_i + (2/3) |\Psi|^6 \tilde\nabla_i T ,  \label{Pconstrain}
\end{equation}
where $W_i$ stands for  the gyrogravitational (or gravitomagnetic) vector potential\cite{MParadigm}; and the operator  $\tilde \bigtriangleup^{\star}$ is defined by\cite{WheelerCIU}
 \begin{equation}
 \tilde\bigtriangleup^{\star} W^{i} \equiv  \tilde\nabla_j ( \tilde \nabla^{i} W^j +  \tilde \nabla^{j} W^i -(2/3)\tilde g^{ij} \tilde\nabla_k W^k). 
\end{equation}
  The role of   the Lichnerowicz  equation (\ref{Lequation})  and the momentum constrain (\ref{Pconstrain}) is the maintenance of general covariance\cite{Dirac}. Remarkably, it is seen that (\ref{Lequation}) is exactly the (GL) equation (\ref{GLequation}) for  superconductivity,  
 and (\ref{Awave}) is similar to (\ref{Pconstrain}). A direct comparison gives some very useful relations. 
 Setting 
\begin{equation}
 \Psi=\sqrt{\rho} \exp(i \mathfrak{e}\theta /2c),
\end{equation}
in Eq.(\ref{Lequation}) we get, collecting the imaginary and real parts of the equation, 
 \begin{equation}
    \,^{(3)}\tilde R \rho + 8 v^2\rho  - \tilde M\rho^{-3} +(2/3)T^2\rho^3 - \tilde Q\rho^{-1} + U \rho =0 \label{WheleerDewitt}
   \end{equation}
   and 
\begin{equation}
\tilde{\nabla}\cdot \rho v  = 0,  \label{continuity}
\end{equation}
where $v$ is the velocity potential and  $U=-(8/ \sqrt{\rho})\tilde\bigtriangleup \sqrt{\rho}$ is related to some sort of compression energy. Take notice that the  ADM-energy formula \cite{ADM} can be used to write:
\begin{equation}
E[g]-E[\tilde g]=-4\oint_\infty  \tilde \nabla |\Psi| \cdot d\tilde \Sigma,
\end{equation}
a generalization of the Newtonian Gauss's law.  It is read directly, from (\ref{GLequation}) and (\ref{Lequation}), that  
 \begin{eqnarray}
 \alpha \propto \,^{(3)}R  \; \; \mathrm{and} \; \; \beta \propto (2/3)T^2|\Psi|^2. \label{bet}
 \end{eqnarray}
 Furthermore,  we shall see shortly that  
 \begin{equation}
 \sqrt{\Lambda/3} \propto -i\mathfrak{e}.
 \end{equation}
Meaning that vacuum energy, which is a source of  gravitating field---hence the extra $|\Psi|^2$ factor in the r.h.s of Eq.(\ref{bet})---performs as a charge. From this perspective the tiny jump in value, from zero to non-zero, of the cosmological constant driving the accelerated expansion of the universe is a symptom of the quantisation of charge.
 
 \section{The principle of least action and the superconductivity of gravitation}\label{Vsection}
 
 The aim of this section is to present how the fundamental laws extending the notion of superconductivity into the realm of gravitation can be put in the form of a principle of least action, clarifying the role played by $\Psi$ and $\Lambda$ in the corresponding superconductivity theory. This exercise demands expressing the scalar curvature in a new way. To do this  
 suppose that $(M, g_{\mu \nu})$ is a globally hyperbolic spacetime foliated by Cauchy surfaces $\Sigma_t$ parameterized by a global time function. Now, instead of writing down the familiar ADM decomposition for the 3+1 splitting of the spacetime\cite{ADM}, consider its `dual' defined by the line element
  \begin{equation}  
d{\mit s}^{2}_{1+3} = g_{\mu \nu} dx^{\mu} dx^{\nu}= -{N}^2 (cdt +A_{l} dx^{l})^2+ {a}^2 \tilde \lambda_{ik}dx^{i}dx^{k}, \label{metric}
\end{equation}
 we shall see that in this way the 3-space vector $A_j$ approximates better the effects of a gravitomagnetic (or  gyrogravitational) potential. Greek indices are used here to indicate four dimensional quantities, whereas Latin indices are reserved to denote three dimensional ones. 
 The conformal transformations symmetries of the 3-space metric  are followed by inserting the scale factor $a(t,\vec{x}).$  $N$ is a redshift function and $c$ the velocity of light.  This form for the line element of the spacetime can be further motivate by an important feature of the equations governing stationary axisymmetric spacetimes, where in that case a transformation between the time and the azimuthal angle coordinate: $t\rightarrow i \varphi$ and $\varphi \rightarrow -it$ leads to a conjugate solution of the same Einstein's equations\cite{Chandra}. 
 
 The components of the metric related to this 3+1 splitting of the spacetime $(M, g_{\mu \nu})$ are given by
\begin{eqnarray}
g_{\mu \nu}=
   \left(\begin{array}{cc}
g_{00} & g_{0k} \\
 g_{i0} &   g_{ik}    
   \end{array}\right)
=   \left(\begin{array}{lc}
-N^2 & -N^2 A_k \\
 -N^2 A_i & {a}^2 \tilde \lambda_{ik} -  N^2 A_i A_k
   \end{array}\right).
\end{eqnarray}
The inverse metric reduces to
\begin{eqnarray}
g^{\mu \nu}=
   \left(\begin{array}{cc}
g^{00} & g^{0m} \\
 g^{k0} &   g^{km}    
   \end{array}\right)
=   \left(\begin{array}{cr}
 a^{-2} A^{l}A_{l}- N^{-2} & - A^{m} /a^{2} \\
   -A^{k}/a^{2}   &    \tilde \lambda^{km}/ {a}^{2}    
   \end{array}\right),
\end{eqnarray}
where
\begin{equation}
\tilde \lambda^{ik} \tilde \lambda_{kj}=\delta^{i}_{j}.
\end{equation}
The indices in $A_i$ and $A^{k}$ are rise and lowered using $\tilde \lambda^{ik}$ and $\tilde \lambda^{kj}$ respectively (unless otherwise indicated). Notice that the role of  $g_{\mu \nu}$ and $g^{\mu \nu}$ have been inverted if a comparison is made  with the ADM setting.
 The induced metric $h_{ij}$ obtained by constraining the  $t-$coordinate to a constant value becomes
\begin{equation}
 h_{ik} = {a}^2 \tilde \lambda_{ik} -  N^2 A_i A_k,    
\end{equation}
(as a side note it is instructive to observe that the Kerr-Schild form of the Kerr-spacetime metric  reduces to $\eta_{\mu \nu}+\ell_\mu \ell_\nu,$ where $\ell_\mu$ is a null vector and $\eta_{\mu \nu}$ is the Minkowski metric\cite{Chandra}). 

The inverse metric of $h_{ij}$ is thus given by
\begin{equation}
 ({h^{-1}})^{ik} = \frac{\tilde \lambda^{ik}}{a^2}+  \frac{\gamma^2}{a^4}N^2 A^i A^k,    
\end{equation}
 where
 \begin{equation}
\gamma\equiv (1-  \frac{N^2 A^{l}A_{l}}{a^2})^{-1/2},
 \end{equation}
might be viewed as a sort of  Lorentz contraction factor.
In effect, a direct calculation shows that:
  \begin{equation}
 ({h^{-1}})^{ik} h_{kj}=\delta^{i}_{j}.
\end{equation}
The peculiar form of the matrix multiplication $ ({h^{-1}})^{ik} \lambda_{kj}$ can be used to deduce the  determinant of $h_{ij},$ which reduces to:
\begin{equation}
(\det h_{ij})^{1/2} = \gamma^{-1}a^{3}(\det \tilde \lambda_{ij})^{1/2}. \label{deth}
\end{equation}
In the system of coordinates given by (\ref{metric}), the  unit normal  $\hat{n}$ to the submanifold $\Sigma_t$  obtained by  making 
the $t$-coordinate equal to a constant is given by 
\begin{eqnarray}
(n^{0},n^{k})=\gamma(1/N, A^k/a^2), \\  (n_{0},n_{k})=\gamma(-1,A_i).
\end{eqnarray}
The trace of extrinsic curvature $K$ of $\Sigma_t$ (i.e. $-g^{\mu \nu} \,^{(4)}\nabla_{\mu} \hat{n}_{\nu}$)  reduces to
\begin{eqnarray}
h^{ij}K_{ij}=  -\frac{\gamma}{2cN} (\dot{a}/a)(\delta^i_i - \frac{\gamma^2 N^2}{a^2} A^{i} A_{i}  ) \\
- 2Na^{-2}(\gamma A^i)_{.i}.  \label{Kcurvature}\nonumber
\end{eqnarray}
The four dimensional Ricci scalar on the other hand can be cast (after some algebraic manipulations) into a sum of familiar terms: including the Ricci scalar for the 3-space metric $\tilde \lambda_{ij}$, a gravitomagnetic field stress action term ($F_{ij}=A_{j,i}-A_{i,j}$) with its characteristic---and curious---sign in front, a  FRLW allotment, a Stueckelberg-Proca piece,  a total derivative term,  and the remnant. Thus we have the following basic relation
\begin{eqnarray}
\,^{(3+1)}R(g)=a^{-2} \,^{(3)}\tilde R+ 4^{-1} N^2  g^{il}   g^{jk}  F_{lk}F_{ij}  \nonumber  \\ \nonumber
+6c^{-2}N^{-2}[ (\dot{a}/{a})^{\cdot} + 2  (\dot{a}/a)^2 ] \\ \nonumber
- 2  a^{-2}g^{ij} (a_{,i} - c^{-1} A_i \dot{a} )( a_{,j} - c^{-1} A_j \dot{a} ) \\ \nonumber
-4   a^{-2}\tilde \nabla^{k}  [a^{-1} (a_{,k} - c^{-1} A_{k}\dot{a}) ] \\ \nonumber
+4 g^{ij} [(c^{-1} A_i  (\dot{a}/a)_{,j}      -c^{-2} A_i A_j (\dot{a}/a)^{\cdot} ],  \label{R4} \\ 
\end{eqnarray}
where $a^{-2}\tilde \lambda^{ij}=g^{ij}$ is the physical metric and $\tilde \nabla^{k}$ denotes the covariant derivative with respect to the $\tilde\lambda$-metric. For simplicity  $N$ is taken constant.  By examining Eq.(\ref{R4}), it is seen that the addition of a cosmological constant (and its relation with an imaginary charge) brings the similarities between gravitation and quantum electrodynamics a little bit closer.  
 The gauge transformation 
 \begin{eqnarray}
\psi(\vec{x}) \mapsto \psi'(\vec{x}) = e^{i\frac{q}{c\hbar}\alpha(\vec{x})}\psi (\vec{x})\\
A_k \mapsto A_k^{'}=A_k-\partial_k \alpha(\vec{x}) 
\end{eqnarray}
leaves invariant the logarithmic derivate \begin{equation}
  D'_k \ln \psi' = D_k \ln \psi
\end{equation}
 where $D_k =\partial_k-(iq A_k /c\hbar).$ 
To gain some intuition let us assume first that
\begin{equation}
a=e^{-\sqrt{\Lambda/3} N c t} \psi(\vec{x}) \label{ALamb}
\end{equation}
in Eq.(\ref{R4}).
Then, a complex structure in (\ref{R4}) can be incorporated by setting
\begin{equation}
\psi(\vec{x})= \rho(\vec{x})e^{i \mathfrak{e} \varphi(\vec{x}) }  \label{adef}
 \end{equation}
 That is, the scale factor splits into a modulus field $\rho(\vec{x})$ and a scalar $\varphi(\vec{x})$: which can be interpreted as a Goldstone boson field, where $\varphi(\vec{x})$ is identified with $\varphi(\vec{x})+ 2\pi /\mathfrak{e}.$ 
 Inserting  (\ref{adef})  in the Einstein-Hilbert action
 \begin{equation}
 S=(2\chi)^{-1}\int(\,^{(4)}R-2\Lambda)(-\det g)^{1/2}, \label{EHaction}
\end{equation} 
and writing
\begin{equation}
N(\Lambda/3)^{1/2}=-i\mathfrak{e}, 
\end{equation}
where
\begin{equation}
\mathfrak{e} \equiv n (q\hbar^{-1}G^{-1/2}c^2)N, \; \;  n\in\mathbb{Z},
\end{equation}
 a principle of least action $S=\int \mathcal{L}_s dt$ is obtained, where
 \begin{eqnarray}
L_s= \frac{1}{2\chi }\int N (\det \tilde \lambda )^{1/2}  \bigg\{ 2\mathfrak{e}^2 \rho \tilde \lambda^{ij} (\varphi_{,i}-A_i) (\varphi_{,j}-A_j)  \nonumber \\
+\,^{(3)}\tilde R \rho -\frac{6\mathfrak{e}^2}{N^2} \rho^3 -4\tilde \nabla^k\rho_{,k}  \nonumber \\ 
+2\rho \tilde\nabla_k \ln| \rho| \tilde\nabla^k  \ln |\rho| + \frac{N^2}{4 \rho} \tilde \lambda^{ik} \tilde \lambda ^{jm} F_{ij} F_{km}   \nonumber\\
 -4i\mathfrak{e}\tilde \nabla^k[\rho (\varphi_{,k}-A_k)]  \bigg\} dV.  \; \; \label{Superlagrangian}
\end{eqnarray}
  The $i\hbar^{-1}$ factor multiplying $\varphi$ or $q$, and consequently the presence of $\Lambda$, has the effect of transforming the classical formula (\ref{EHaction}) into a quantum mechanical expression (\ref{Superlagrangian}), where several useful parameters---describing the superconducting state of the four-dimensional spacetime---can be worked out. 
  
  Varying $\rho$ in $\mathcal{L}_s,$ the Lichnerowicz equation
  \begin{eqnarray}
 -8\tilde \Delta |\Psi| + 2\mathfrak{e}^2 (\vec{\tilde \nabla}\varphi-\vec{A})^2 |\Psi|  + \,^{(3)}\tilde R|\Psi |  \nonumber \\ 
 - \frac{18\mathfrak{e}^2}{N^2} |\Psi|^5  - \frac{N^2}{2}\tilde H^2|\Psi|^{-3} = 0, \label{Schr}
\end{eqnarray}
   is recovered, where 
\begin{equation}
|\Psi|=\sqrt{\rho} 
\end{equation}
and the relation $\tilde F^{km} \tilde F_{km}=2\tilde H^2$ has been used. A comparison with  Eq. (\ref{Lequation}) gives the contributions to the conformal density of gravitational-wave effective kinetic energy $\tilde{M},$ the York time $T,$ and the conformal local energy density of ordinary mass energy $Q,$ respectively: 
\begin{eqnarray}
 |\Psi|^{-8}\tilde{M}&=& 2\Lambda N^2 (\tilde \nabla\varphi-\mathbf{A})^2/3, \label{GravitatingMass} \\
 2T^2/3&=& 6\Lambda,  \label{TL} \\
\tilde Q&=&N^2  {\tilde H}^2/2,
\end{eqnarray}
where we have put the cosmological constant $\Lambda$ back. 
   On the light of expression (\ref{GravitatingMass}), it is worth pointing out that initial data sets for energy densities and currents are scaled as 
\begin{eqnarray}
\rho&=& \Psi^{-8} \tilde \rho  \\ 
\mathbf{j}&=& \Psi^{-10} \mathbf{\tilde j}
\end{eqnarray}
 respectively, among other things to preserve the dominant energy condition\cite{WheelerCIU, HawkingEllis}: which implies that\cite{Carter}, ``at the classical level, the vacuum must be stable against spontaneous matter creation process.''   
 
 From $ \mathcal{L}_s $ it is readily seen that if the gravitomagnetic field is a pure gauge, {\it the $U(1)$ gauge symmetry becomes spontaneously  broken} when 
\begin{equation}
\rho=\rho_s=|\Psi_s|^2, 
\end{equation} 
 where
 \begin{eqnarray}
|\Psi_s|^4=N^2 \,^{(3)}\tilde R_s/18{|\mathfrak{e}^\dag|}^2= \,^{(3)}\tilde R_s/6\Lambda.
 \end{eqnarray}
  The last condition characterizes the superconducting state of gravitation and requires $\Lambda\neq 0$.  
  
  The continuity equation 
\begin{equation}
\tilde \nabla^k \tilde {J}_k=0
\end{equation}
 is obtained by the variation of $\varphi$ in $\mathcal{L}_s,$ giving
 \begin{equation}
    \mathbf{\tilde J}=-4 N {\mathfrak{e}^\dagger}^2 |\Psi|^2 (\mathbf{\tilde \nabla}\varphi-\mathbf{A})/2\chi;
\end{equation} 
 that is, 
 \begin{equation}
\mathbf{{\tilde J}}=-\frac{4\mathfrak{e} N}{ 2\chi}[ \Psi (\frac{\hbar}{i} \mathbf{\tilde \nabla}-\frac{\mathfrak{e}}{2}\mathbf{A})^{\ast} \Psi^{\ast} +  \Psi^\ast (\frac{\hbar}{i }\mathbf{\tilde \nabla}-\frac{\mathfrak{e}}{2}\mathbf{A})\Psi]. 
 \end{equation}
 The significance of this is that the superconducting current  $\mathbf{{\tilde J}}$ might be interpreted as a probability flux of Cooper pairs charged with vacuum energy and moving with velocity proportional to $\mathbf{\tilde \nabla}\varphi-\mathbf{A}:$ Perhaps, the picture of a  space-time superfluid charged with vacuum energy might provide a  basis for understanding the exotic phenomena observed in galactic cluster collisions\cite{Clowe}, and in particular the ringlike dark matter structure  observed in the galaxy cluster C1 $0024+17$ \cite{Jee}.
 
  The dynamical momenta $-i \hbar \tilde \nabla \mathfrak{e}\varphi$  does not change suddenly when a gravitomagnetic vector potential is switched on \cite{FeynmanIII}.
  
   Varying $\mathbf{A}$ in $\mathcal{L}_s$ yields 
 \begin{equation}
 \tilde \nabla_k \tilde F^{k}_l=4\frac{{\mathfrak{e}}^2 \rho^2 }{N^2}(\tilde \nabla_l \varphi-A_l ).  
 \end{equation}

The inverse square root  of the coefficient on the r.h.s of the previous relation gives  the penetration depth
 \begin{equation}
 \lambda_s= 3 (2\, \,^{(3)} \tilde R_s)^{-1/2}. \label{Plenght}
 \end{equation}
 This number measures how the gravitomagnetic field decays with distance deep inside a large spacetime-superconducting zone---say by a sort of gravitomagnetic Meissner effect\cite{Feynman1972, WeinbergQFT}; it also determines the thickness of the surface layer where the superconducting current can flow.  Expressing the Lichnerowicz equation in the form  
 \begin{equation}
 \tilde \Delta |\Psi|=\tilde P(|\Psi|),
 \end{equation}
  and  evaluating  $\partial \tilde P(|\Psi|)/\partial {|\Psi|}$ at $|\Psi|=|\Psi_s|,$ the correlation length $\xi_s$ can be obtained
 ($-\tilde \triangle \delta |\Psi|=\xi^{-2} \delta|\Psi|$).  It reduces to 
 \begin{equation}
\xi_s=\sqrt{2} \, \,^{(3)} \tilde R_s^{-1/2}. \label{Colength}
 \end{equation}
   \begin{figure}[t]
\begin{center}
\includegraphics[width=3in]{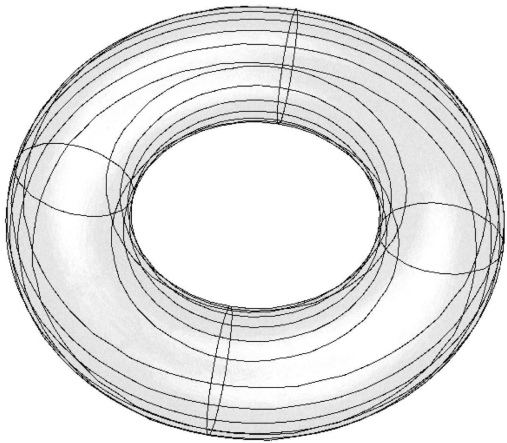}
\caption{Superconducting currents on a torus surface following circular orbits. A pair of cross sections of the torus have also been depicted. In the interior, the modulus of the gravitational potential $|\Psi|$ is kept nearly constant. The thickness of the surface layer where the superconducting current can flow is controlled by the value of penetration depth $\lambda_s= 3 (2\, \,^{(3)} \tilde R_s)^{-1/2}.$  The extension of the Cooper pairs are of the order of  $\xi_s=\sqrt{2} \, \,\,^{(3)} \tilde R_s^{-1/2}.$ } \label{Torus}
\end{center}
\end{figure}
  This number measures the extension of the Cooper pair, and it is also the distance through which a change  will spread  if a small fluctuation of the superconductor state occurs at a given point.  
  
  The energy gap  per unit volume $\triangle_{N-S}$ between the normal state ($\rho=0$) and the superconductor state ($\rho=\rho_s$) can be obtained by evaluating the integrand of $\mathcal{L}_s$  \cite{WeinbergQFT},    giving 
 \begin{equation} 
    \triangle_{N-S}= 2^{1/2}9^{-1}{\mathfrak{e}}^{-1} N \,^{(3)}\tilde R_s^{3/2}.
      \end{equation}
  This energy gap is related to a critical magnitude $\tilde H_c,$ of the gravitomagnetic field which when exceeded drives the spacetime to its normal state; i.e. when the energy cost per volume to expelled the gravitomagnetic field $N^2 {\tilde H}^2/2$ is greater than $\triangle_{N-S}.$  This leads (for a  sufficiently large superconducting  zone compared with $\xi$) the value 
  \begin{equation}
 \tilde H_c= N^{-1}\sqrt{2\rho_s\triangle_{ N-S}}=3^{-3/2}2^{1/2}|{\mathfrak{e}}^{-1}  \,^{(3)} \tilde R_s|.
 \end{equation} 
   It follows that the superconductivity of gravitation is destroyed in a region when there is a sufficiently strong gravitational spatial curvature. Hence, the most probable place to observe the superconducting phenomenon just described is in the outer skirts of galaxies and not near a central region where a supermassive black hole might be present.  Inserting the above values (\ref{Plenght}) and (\ref{Colength}) in the dimensionless Gizburg-Landau parameter defined by $\kappa \equiv \lambda_s/\xi_s,$
 it is discovered that the superconducting state of gravitation is of the second kind (or type II \cite{Abrikosov, WeinbergQFT}) as   
   \begin{equation}
     \kappa= 1.5
   \end{equation}
is bigger than one. Under this classification also fall other substances like niobium, heavy fermionic materials, fullerenes, and high-temperatures superconductors. It is predicted; therefore,  that quantum gravitational effects in more severe circumstances may generate stable vortex lines of minimum flux on the fabric of the spacetime, precisely  when  the strength for an external gravitomagnetic field lies between $\tilde H_{c1}\sim\kappa^{-1} \tilde H_c$ and  $\tilde H_{c2}\sim\kappa \tilde H_c$ (the Shubnikov phase), which might have important implications for cosmology.

 The imaginary part of the Lagrangian is a total derivative and it does not affect the equations of motion. Natural boundary conditions, such as the vanishing of the normal component of the current $\mathbf{\tilde J}$  at the boundary surface with unit normal $\mathbf{\hat{n}},$ 
\begin{equation}
\mathbf{\tilde J} \cdot \mathbf{\hat{n}} = 0,
\end{equation}
 can be found by considering the surface integrals appearing in the variation of the action principle (\ref{Superlagrangian}), 
 see also FIG.\ref{Torus}.  
 \section{The quantization of York's time}\label{Ysection}

 Let us stress that by Eq. (\ref{TL}) the quatization of York's time is related to the quantization of the cosmological constant 
 $\Lambda.$  The spirit of this section is to provoke some thought about the relation between quantum mechanics and general relativity. Since we shall deal with cavities, charges, processes where all the fast things have happened and all the slow things not, and  the notion of temperature associated to cosmological horizons,  it will be instructive to note that all these elements form part of the standard setting for the derivation of Planck's radiation formula: A hot cavity containing radiation in thermal equilibrium. Let us start with a very profound question.
   \subsection{``How come the Quantum?''} \label{HowCome}
``Of all the obstacles to understand the foundations of physics,'' John Wheeler used to say \cite{ WheelerMisner}, ``it is difficult to point one more challenging than the question: How come the Quantum?''   To at least start scratching the surface of this mystery we shall consider the Hamilton-Jacobi theory of the York--Time action 
\begin{equation} 
 S_K =  \frac{1}{\chi}  \int_{t=\Phi^{+}(\vec{x})} K d\Sigma  - \frac{1}{\chi}  \int_{t=\Phi^-(\vec{x})} K d\Sigma
 \label{YorkAction}
\end{equation}
so that we can make more easily a leap from classical ideas to quantum ones. Let us focus on the case of a thin-sandwich-spacetime system. Let the interior of this spacetime configuration (or hot cavity) be given by a strip of de-Sitter spacetime ($int \, M=dS_4$) whose metric can be cast into the form:
\begin{equation}
ds^2=-c^2d\tau^2+ e^{-2(\Lambda/3)^{1/2}c\tau} \delta_{ik}dx^{i}dx^{k},
\end{equation}
which is the line element of the steady-state universe of Bondi, Gold, and Hoyle \cite{Birrell}.
Let the top (bottom)  be a $3$-space hypersurface of the form 
$\Sigma_{t=\Phi^+ (\vec{x})}$ $(\Sigma_{t=\Phi- (\vec{x})}).$ Thus, from the four dimensional perspective, by  (\ref{metric})  and (\ref{ALamb}), $A_j$  becomes pure gauge in the neighbourhood of each of these hypersurfaces, say $A_j=c\nabla_j \Phi^+$ at the top and $A_j=c\nabla_j \Phi^-$ at the bottom. What is important is how this quantities are related. Let us introduce a new set variables: The time interval
\begin{equation} 
\theta =N (\Phi^+(\vec{x}) - \Phi^-(\vec{x}) ) \label{Trick}
\end{equation}
and the {\it relative scale factor} $\varrho$ defined by
\begin{equation}
\frac{1}{4}\varrho^2= \sqrt{\Lambda/3} \chi^{-1} ( e^{(-\sqrt{\Lambda/3} Nc\Phi^-(\vec{x}))}-e^{(-\sqrt{\Lambda/3} Nc\Phi^+(\vec{x}))} ) \label{aRel}
\end{equation}
Inserting (\ref{deth}) and (\ref{Kcurvature}) in (\ref{YorkAction}), with $a= e^{-Nc\sqrt{\Lambda/3} \Phi(\vec{x})}$, it is found that $S_K,$ up to second order of approximation in the derivatives of  $\theta$ and $\varrho,$ reduces to a simple---but remarkable expression, that rests entirely on the fundamental principles of the general theory of relativity:
 \begin{eqnarray}
 S_K = \int_{\mathbb{E}^3} \delta^{ij} ( \frac{1}{4}  Z^2\varrho^2 \vec{\nabla}_i\theta \vec{\nabla}_j\theta - \vec{\nabla}_i \varrho \vec{\nabla}_j \varrho)  \nonumber   \label{PlanckRad} \\ + \mbox{\it higher order terms.}  \label{Superposition}
\end{eqnarray} 
where $Z$ is given by:
\begin{equation}
Z=\frac{e^{\mathcal{J} \omega^{\ast}/2K_B T}}{ e^{\mathcal{J} \omega^{\ast}/K_B T}-1}=\sum_n e^{\mathcal{J} \omega^{\ast} (n+\frac{1}{2})/K_B T} \label{Planckian}
\end{equation}
Notice the conspicuous similarity of $Z$ with the partition function of an ideal Bose Einstein gas (``quantum mechanics without quantum mechanics?''). The following identifications have been made however:
\begin{eqnarray}
K_BT&=&\frac{\hbar c}{2\pi} \sqrt{\Lambda/3}, \\
\sqrt{\Lambda/3}  &=& -i\frac{ e^\dagger}{c\hbar} \sqrt{ 1/4\pi \epsilon_oG} c^2, \label{Cosmo}\\
\mathcal{J}&=&\frac{{e^{\dagger}}^2}{c} \label{Jaction},
\end{eqnarray}
and
\begin{equation}
\omega^*= t_{P}^{-1} (c \theta/2\pi \ell_{P}) \label{Ytime}.
\end{equation}
This first relation comes from associated  temperature of the cosmological horizon. The second relates $\Lambda$ with the charge 
$e^\dagger.$  $\Lambda$ itself can also  be used to define a natural unit of action $\mathcal{J},$  as in the electrogravitic scale\cite{Visser}.  Thus, when $e^\dagger,$ lets say, is made by hand numerically equal to the charge of the electron, $J$ reduces to $\hbar$ times the fine structure constant $\alpha\approx 1/137$. $t_{P}$ and $\ell_{P}$ are the Planck time and Planck length respectively. 

We shall now proceed heuristically, but later on we will provide another argument leading to similar findings. 
 We might associate $\omega^*$ with a discrete portion of energy as given by  
\begin{equation}
 E_{n_+}-E_{n_-}=\mathcal{J}  \omega^{\ast}, \label{Bohrsfrequency}
\end{equation}
which can be viewed as an extension of  Bohr's  frequency condition.  In view of Eq. (\ref{Ytime}), let us  take the natural step forward of assuming that the time difference $\theta$ is quantized: in such a way that, in the proposed period, a light ray would girdle an integer number of times as if it were to trace a flat circle of radius $\ell_{P},$ and explore the consequence of this in a little more general situation: say when  $H=\nabla \times A$ doesn't hold globally, which could be the case if we add to the physical system a gravitomagnetic monopole or other type of topological obstruction. For instance, one might assume a `Dirac  string' ending in some point inside our spacetime sandwich. Take a 2-sphere immersed in $ I \times \mathbb{E}_2,$ where $I$ represents a sufficiently long interval of time; then   
\begin{eqnarray}
\Phi_{flux}  &=& \int _\odot \nabla \times  A_{+} d\Omega_+ -  \int _\oplus   \nabla \times  A_{-} d\Omega_-  \label{BSommer}\\ \nonumber
&=&\oint c\vec{\nabla}\theta \cdot d\vec{s}  \\ \nonumber
&=&  2\pi \ell_P n;   \; \;    n \in \mathbb{Z}.
 \end{eqnarray}
The first line correspond  to the flux over a close sphere which has been divided into upper and bottom hemispheres. Using Stokes's theorem the added integrals of the first line are converted into a  single close path integral over the equator.  $A_{+}$ and $A_{-}$ are   related by means of a gauge transformation. Using our above result about the quantization of time one arrives at the third line.   
Which states the quantization of the vortex strength. The smallest vortex has circulation  $2\pi \ell_P.$  The last two lines in Eq.(\ref{BSommer}) can be regarded  as a  Bohr-Sommerfield quantization condition on a completely accessible close path\cite{Fermi}  
\begin{equation}
\oint p dx = 2\pi \hbar n.
\end{equation}  
 Alternatively, from (\ref{Cosmo}), (\ref{Jaction}), and (\ref{Ytime}),  we might claim that it is  the cosmological constant the one that has been quantized  and given in terms of multiplets of a fundamental unit of charge: $-i\mathfrak{e}$. 

The first term in (\ref{PlanckRad}) can be regarded as a generalization of the relation 
\begin{equation}
 E_G =   (4G)^{-1} \int_{\mathbb{E}^3}  (\vec{\nabla} \Phi^+(\vec{x}) - \vec{\nabla} \Phi^-(\vec{x})   )^2, \label{TimeCollapse}
 \end{equation}
since Eq. (\ref{PlanckRad}) includes an extra weighting  factor given by $\varrho^2 Z^2.$ Eq. (\ref{TimeCollapse}) was introduced  by Penrose in the context of the gravitational reduction of the wave packet\cite{Penrose, Anandan}. Let us see if we can gain some insight from this circumstance. In Penrose's proposal a lump of mass is placed in an unstable superposition state $(1/\sqrt{2})(|-> +   \; |+>).$ The states: `here' $|->$  and  `there' $|+>$  form a preferred basis of states, mysteriously  chosen  by nature itself (an affair known as the preferred basis problem).   Let $\Phi^+(\vec{x})$ and  $\Phi^-(\vec{x})$  be the gravitational potential energy for each of the above states respectively. Notice that  in (\ref{Trick}), (\ref{TimeCollapse}),  and therefore in (\ref{Superposition}), there is a delicate issue for making a pointwise identification of two different spaces since one must ensure that the principle of general covariance is preserved. By  instability we mean  that  the linear unitary evolution of states  $(1/\sqrt{2})(|-> + \; 
 |+>)$  is not going to last. 
 ``Schr\"odinger's equation'' quoting John Bell\cite{Bell, Ghirardi} ``is not always right.''   Then, according to the wave packet reduction scenario, a pure unitary evolution is only an approximation, and another piece of the quantum mechanical setting sets foot in; namely, a non deterministic, time asymmetrical (non-local) law of evolution that makes the wave function `collapse' into $|->$ OR  $|+>$. Since mass affects the rate of clocks, a fuzziness in the description of time is manifest from the very beginning by the consideration of  such linear superposition of states, a blurriness scaling all the way down to  the Schr\"odinger equation itself. Following the analogy with other quantum unstable systems, Penrose argued that  $E_G$ is set to capture not only the indefiniteness of the energy of the transient state $(1/\sqrt{2})(|-> + \; |+>)$ but also its lifetime\cite{PenroseH}:
\begin{equation}
T_G\sim \frac{1}{E_G}. \label{decay}
\end{equation}
This gives physical meaning to  (\ref{TimeCollapse}) and (\ref{Superposition}).
A  central issue that needs to be tackled, however, is the law conservation of energy, one of  the cornerstones in physics. Pioneering work on the subject of (objective) wave packet reduction have ran into trouble with this law\cite{GRW}. But it has been anticipated that bringing gravity into the picture might fix the problem (no matter how tiny it is). To test 
the prediction of the rate of state reduction given by  Eq.(\ref{decay}),  and specially to explore whether or not  the phenomenon of wavefunction collapse is a real physical process, Space and Earth base experiments have been proposed\cite{Marshall, Wezel}.  We have not succeed yet in providing a dynamical formulation  for Penrose proposal, but perhaps the ideas set forth here can lead,  under further investigation, to a novel dynamical formulation for the gravitational collapse of the wave function. In ref.\cite{WezelSymmetry}, it is argued that time translation symmetry can be spontaneously  broken in such a way that the Schr\"{o}dinger's equation becomes perturbed infinitesimally by a weak unitarity breaking field, where the perturbation is assumed to be due to the influence of general relativity.  

Let us turn now to another argument, and see what happens when we in fact  invoke quantum mechanics to study the equations of  motion  derived from (\ref{YorkAction}). Are we led to similar conclusions? 
\subsection{The instability of Misner space \\ and the fall of a particle to the centre}\label{MisnerSection} 
 Introducing the variable 
\begin{equation} 
\varphi=  2^{-1} \int Z d\theta= 2^{-1} \ln |\tanh (4^{-1}\theta)|,
\end{equation}
the metric
 \begin{eqnarray}
ds^2=- d\varrho^2 + \varrho^2 d^2\varphi,  \label{Mmetric}
\end{eqnarray}
can be read off from equation (\ref{PlanckRad}). It is the metric of  the Misner space; that is, Minkowski space with identification under a boost \cite{Thorne}.
 \begin{figure}[t]
\begin{center}
\includegraphics[width=1.8in]{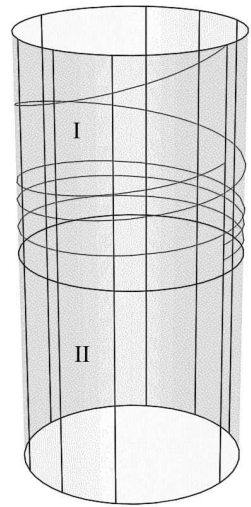}
\caption{A locally inextendible analytic extension of Misner space with topology $\mathcal{R}\times I$. The time $t$ runs along the vertical axis.  The chronology horizon, 
at $t=0,$ is depicted by the circle at the centre of the diagram and  separates region I from a causality violating zone: region II.  The vertical null geodesics $\phi'=cte.$ are complete, in contrast the twisted null geodesics, pirouetting an infinite amount of times near the chronology horizon, are incomplete.} \label{Myspace}
\end{center}
\end{figure}
Misner space is a  geodesically {\it incomplete} spacetime  with topology $\mathcal{R}\times S^1;$ it contains close time-like curves (CTC's), and a chronological horizon at the critical value $\varrho=0,$  see Fig.\ref{Myspace}. Changing  variables   ($\varrho \rightarrow t^{1/2};$  $\varphi \rightarrow 2^{-1}\phi),$ we get: 
\begin{eqnarray}
ds^2=4^{-1}(- t^{-1} dt^2 + t d^2\phi). \label{tmetric}
\end{eqnarray}
Setting $\phi^{\prime}=\phi-\ln t$ in (\ref{tmetric}), the metric reduces to:
\begin{equation}
ds^2= 2dt d\phi^{\prime} + t d^2\phi^{\prime}, \label{round}
\end{equation}
which is non singular at $t=0.$ From (\ref{Mmetric}) the geodesic equations can be cast as  
\begin{eqnarray}
\ddot{\varrho}=-\varrho \dot{\varphi};  \label{falling} \\
(\varrho^2 \dot{\varphi})\dot{\,}=0.
\end{eqnarray}
The second equation can be interpreted as the law of conservation of angular momentum; that is
\begin{equation}
 \varrho^2 \dot{\varphi}= \ell=cte. \label{angular}
 \end{equation}
  If $\ell=0,$ either $\rho=0$ (for finite $\dot{\varphi}$) or $\varphi=cte$ (in which case $\varrho= \beta \tau + \varrho_o).$  If $\ell \neq 0,$ writing $\varrho=1/U$ and $d/dt=\ell \varrho^{-2}d/d {\varphi},$ Eq.(\ref{falling}) reduces to $U_{,\varphi\varphi}=U,$ leading to following solutions:
\begin{eqnarray}
\varrho^{-1}&=&\varrho_{\max}^{-1} \cosh(\varphi-\varphi_o);  \label{rmax} \\
\varrho^{-1}&=&\mathcal{C}e^{\pm (\varphi-\varphi_o)};    \label{nullgeo}\\
\varrho^{-1}&=& \sqrt{2\epsilon} \ell^{-1}\sinh(\varphi-\varphi_o) 
\end{eqnarray}
for timelike, null, and spacelike geodesics respectively; $\epsilon$ is  a constant parameter that can be interpreted as a sort of  energy. It is negative for the bound states given by (\ref{rmax}) and it is defined by 
\begin{equation}
\epsilon= 2^{-1} (\dot{\varrho}^2 - \varrho^2 \dot{\varphi}^2). 
\end{equation}  
From (\ref{nullgeo}) it is seen that null geodesics spiral  round and round as they approach to the locus of points satisfying $\varrho=0.$ They are divided symmetrically  into two families,  according to the sign of the exponent.  In the coordinate system given by (\ref{round}),  the null geodesics of one of the families have been untwisted so that they become vertical lines that cross  the chronology horizon at $\varrho=0$; meanwhile the geodesics of other family cannot be analytically continued beyond the horizon, and have {\it  finite} affine length\cite{HawkingEllis, Thorne}.  A symmetric construction can be done by setting  $\phi^{\prime}=\phi-\ln t$ instead, where
 the roles played by both families of null geodesics become  interchanged. This gives  two inequivalent, locally inextendible, analytic extensions which are geodesically incomplete,  see Fig.\ref{Myspace}.   

A test particle which classically would  follow the timelike  geodesics  (\ref{rmax}), and therefore it  is confined to $\varrho \leqslant \varrho_{\max},$  can explore larger values than $\varrho_{\max}$ by quantum tunneling. 

Using (\ref{angular}) we can eliminate $\varphi$ from (\ref{falling}) to get an inverse quadratic potential (or an inverse cube forced), which is singular at the origin ($\varrho=0$), i.e. at the chronology horizon:
\begin{eqnarray}
\ddot{\varrho}=-\partial_{\varrho}V;  \, \, V(\varrho)=-2^{-1}\ell^2 \varrho^{-2}.
\end{eqnarray}
The corresponding Hamiltonian is given by
\begin{eqnarray}
\hat{\mathcal{H}}\Psi(\varrho)=2^{-1}[(-i\hbar \partial_{\varrho})^{2}  - \ell^2 \varrho^{-2}] \Psi(\varrho)=\epsilon \Psi(\varrho). \label{Quanta}
\end{eqnarray}
From the point of view of spectral theory, an inverse quadratic singular potential (or cubic force) is special, in the sense that it does not belong to the Kato's class, and it cannot be regarded as a lower order perturbation of the Laplacian: it marks the division of the appearance of  unusual spectral behaviour not present in less singular potentials (like the ones that at leading order show a power law  dependence in $\varrho$ near the origin of the form $V\sim \mathcal{C}'\varrho^s$, $s>-2$). Notice that for a non-relativistic particle trapped inside a spherical shell of radius $\varrho,$  the Heisenberg's  uncertainty principle leads to an uncertainty in its kinetic energy ($K.E.$) of the order of  $ K.E. \sim \hbar^2 / 2\mu \rho^{2};$ thus, in the form of an inverse square law. Potentials with a roughly inverse-squared-law type behaviour can be found in the Efimov effect\cite{Efimov}, in dipole-electron system\cite{Camblong},  in the near-horizon physics of some black holes\cite{Claus}, and in quantum chromodynamics (QCD); 
for instance,  a na\"ive  perturbative analysis with resummed self energy bubbles for the gluon propagator yields (according to thermal field theory\cite{Bellac}) the following potential for gauge invariant sources: $V(\varrho)\propto \varrho^{-2} e^{-2m_D\varrho},$ where $m_D$ is the Debye mass given by $m_D^2\simeq 3^{-1}(N+N_f)g^2(T)T^2;$ $N,$ $N_f,$  $g(T),$ and $T$ being respectively the number of colours, the number of  flavours, the gauge coupling constant, and the temperature:  $N^{QCD}=3,$ $N^{QCD}=6,$  and $g\sim 1/\ln (T/\Lambda_{QCD}).$    

For a bound stationary state set
\begin{equation}
\epsilon= -\kappa^2. \label{bound}
\end{equation} 
Then,  it is known that if  the constant parameter $\ell$ (with units of action) is larger than some critical value, the spectrum of  (\ref{Quanta}) is {\it continuous.} That is, if 
\begin{equation}
\ell>\hbar/2,
\end{equation}
 no matter what negative is the value of the energy we choose, a quadratically integrable, continuos, wave function which is finite  at infinity, and satisfying    (\ref{Quanta}), can be found \cite{Morse, Landau}. In contrast, in the hydrogen atom one obtains the discrete spectrum found by Niels Bohr.
 
  Remarkably, as it was noticed in \cite{Morse}, a bizarre quantization rule is obtained if one imposes the further requirement that the state functions for bound states be mutually orthogonal: This does not uniquely fix the energy levels,  it 
 fixes, however, the energy levels relative to each other as follows\cite{Morse} 
\begin{eqnarray}
\epsilon_n= -\kappa_o^2 e^{-\frac{2\pi n}{p}}; \; \; \; n=\hdots  , -2,-1, 0,1,2,3, \hdots \label{StrangeQuantization}
\end{eqnarray}
An accumulation point sets in for $n\rightarrow +\infty;$ that is,  near $\epsilon=0.$ 
To see all this more closely  observe, using (\ref{bound}), that a bound-stationary state will obey the  Schr\"odinger's equation 
\begin{equation}
\frac{d^2R}{d\varrho^2}+[-\kappa^2 + \frac{\gamma^2}{\varrho^2}] R=0, \label{Sch}
\end{equation}
 where $\gamma^2=\ell^2 \hbar^{-2}.$ That the spherical Hankel function of imaginary argument and complex order  
\begin{eqnarray}
\frac{1}{\varrho}R =h_{ip-\frac{1}{2}}(i \sqrt{-\epsilon}\varrho) 
\end{eqnarray}
satisfies  (\ref{Sch}) while not diverging at infinity, where $$p=\sqrt{\gamma^2-(1/4)}$$ and 
$$h_{ip-\frac{1}{2}}(i \sqrt{-\epsilon}\varrho)   =$$
\begin{eqnarray}
\sqrt{\frac{\pi}{2i  (-\epsilon)^{\frac{1}{2}} \varrho}}[\frac{e^{\pi p}}{\sinh (\pi p)} J_{ip}(i\sqrt{-\epsilon}\varrho)  
 - \frac{1}{\sinh (\pi p)} J_{-ip}(i\sqrt{-\epsilon}\varrho) ].  &\nonumber \\
\end{eqnarray}
$ J_{ip}$ and $J_{-ip}$ are Bessel functions with the following asymptotic rules at infinity\cite{Lebedev}:
\begin{eqnarray}
h_{ip-\frac{1}{2}}(i \sqrt{-\epsilon}\varrho) \rightarrow \frac{1}{i  (-\epsilon)^{\frac{1}{2}} \varrho}   e^{-\sqrt{-\epsilon} \varrho-\frac{1}{2}i\pi(i p+\frac{1}{2})}, \,\,  \sqrt{-\epsilon}\varrho
 \rightarrow \infty;  \nonumber \\
\end{eqnarray}
and at the origin of coordinates:
\begin{eqnarray}
& h_{ip-\frac{1}{2}}(i \sqrt{-\epsilon}\varrho)  \rightarrow   \sqrt{\frac{2i  \pi}{(-\epsilon)^{\frac{1}{2}} \varrho}} \frac{e^{\frac{1}{2}\pi p}}{ | \Gamma(1+i p) | \sinh (\pi p)} 
 \times \nonumber \\
& \times 
 \sin[p \ln (\frac{1}{2}  (-\epsilon)^{\frac{1}{2}} \varrho ) -  \Theta_p ];  
   \,\,  \sqrt{-\epsilon}\varrho  \rightarrow 0,
\end{eqnarray}
where  $ \Gamma(1+i p) =| \Gamma(1+i p) | e^{i\Theta_p}.$ 

Precisely for $p$ real (i.e. $\gamma>1/2$), no matter how negative is $\epsilon_n,$ the wave function remains finite but oscillates  without limit as $\varrho$ goes to zero. One therefore concludes that the `normal state' corresponds to  $\epsilon_n \rightarrow -\infty,$ where the particle becomes confined to a infinitely small region near the origin $\varrho=0;$ hence, it falls to the centre\cite{Landau}. The scalar product between two of the above eigenfunctions is given by
\begin{equation}
(\kappa_i - \kappa_j )\int_0^\infty R_iR_j= \frac{2i\pi p}{\sqrt{\kappa_i \kappa_j }} \frac{e^{\pi p}  \sin[p\ln |\kappa_i /\kappa_j |]  }{ | \Gamma(1+i p) |^2 \sinh^2 (\pi p)} 
\end{equation}
  which is zero for $i\neq j$ if  $p\ln |\kappa_i /\kappa_j |=n\pi,$ $n\in \mathbb{Z}.$ 
  
It follows from (\ref{Quanta}), the conservation of energy, and (\ref{StrangeQuantization}) that at the turning point 
  \begin{equation}
\varrho^{\max}=\varrho_o^{\max}e^{\frac{\pi n}{p}}; \; \; n \in \mathbb{Z}. 
\end{equation}
Using (\ref{aRel}), fixing $\Phi_+,$ and going to the large $n$ limit it is inferred that
\begin{equation}
e^{\frac{2 \pi n}{p}} \approx e^{\sqrt{\Lambda/3} N c (\theta-\theta_o) }.
\end{equation}
where $ e^{-\sqrt{\Lambda/3} N c \theta_o } \propto e^{-\sqrt{\Lambda/3} Nc\Phi_+}$. 

Hence,  in the proximity of the accumulation point, where $\epsilon$ vanishes, it is found that
\begin{equation}
\sqrt{\Lambda/3}  N c (\theta-\theta_o)\approx 2\pi n/p,
\end{equation}
 meaning that $\omega^{\ast}$ in (\ref{Ytime}) and (\ref{Bohrsfrequency}) is proportional to $n \in \mathbb{Z},$ as was naively presume in section \ref{HowCome}. Alternatively, by (\ref{Jaction}) and (\ref{Cosmo}), this result can also be regarded as a subtle quantization of charge; hence, of $\sqrt{\Lambda/3}.$ The classical instability of Misner space at the chronology horizon  signifies that for any physically reasonable perturbation the spacetime geometry will be radically altered; the complete detail of the transformation and the way it can be related to the wave function collapse scenario, however, remains an open question.       

\section{Remarks and Conclusions}\label{ConclusionSection}
The special theory of relativity made the `luminiferous ether' of Huygens a superfluous entity inasmuch as there is no need to appeal for an absolute stationary space in which electromagnetic waves propagate at the absolute speed $c.$ Thus,  ``it removed from the ether its last mechanical quality, its immobility''. There was no ether: an incompressible, extremely dense, extremely elastic substance that offered no resistance to the passage of matter to it.  Today we are not free from difficulties and we have incorporated an invisible substance---the enigmatic cold dark matter---in our theories, which  has also some striking properties in order to explain some aspects of the universe at large scales.
When cold dark matter is not invoked, theorists resort to modifications of the law of gravity or an alteration of the Newtonian dynamics.  
 The random fluctuations of the quantum world seem to introduce, however, a new class of ether: `the quantum vacuum'  whose physical properties can be determined by studying the way it reacts under external stimuli\cite{DeWitt}. We have argued that it is this feature of the natural world that determines the nature of cold dark matter, and that there is no need to recur to supersymmetric particles, or to the axion, or to  abandon the  Einstein's field equations at large scales; rather, it is proposed that a notion of superconductivity in the realm of gravity is the key to solve this conundrum, providing also a new context to envision the cosmological constant problem: where vacuum energy plays the role of charge. 
 
  According to Sakharov \cite{Sakharov1967}, who based his arguments on previous work by Zel'dovich \cite{Zel1967} on the analysis of the physics of the cosmological constant: 
 
``Gravitation may be regarded as the metric elasticity of space that arises from elementary particle physics.''

The `quantum vacuum' associated with the various fields and particles must react in a precise manner to changes in the curving of space (or to changes in the boundary conditions). Thus, expressing first the  vacuum action  as a sum of terms ordered by its degree of nonlinearity in the curvature for a given geometry (with appropriated boundary terms); and secondly, using  Sakharov's  hypothesis to establish the equivalence  between, the linear term in the curvature of such an expansion and the Einstein-Hilbert action: It is concluded that the Newtonian constant of gravity $G$ is a kind of  `elastic constant of the metric'  whose value is completely determined by  elementary particle physics and a natural cut off scale\cite{MTW}. Formally it is obtained that $G=c^3/(16 \pi A \hbar \int k dk)$ where $A$ is a dimensionless factor of order one, and the divergent integral is over the momenta of the virtual particles.  A rough of estimate of $G$ sends the cut off scale to the Planckian regime
\begin{equation}
k_{\rm cut \, off} \sim (c^3/\hbar G)^{1/2} = l^{-1}_{pl}\approx 1/1.6\times 10^{-33} cm, \label{cutoff},
\end{equation}
 marking the limit of applicability of the theory of quantum fields. If this intuitive insight turned out to be correct, it would mean---by comparison with the theory of elasticity---that: Einstein's gravity is not as fundamental as one would have been expected, being only an emergent aspect of particle physics. The principle of least action (\ref{Superlagrangian}) inferred in section V strongly support this atomistic thinking. 
 
  It must also be admitted, by Eq.(\ref{cutoff}), that it would not be a priori  justified a direct use of one (or perhaps all) of the usual notions of field, particle, space, quantum, or time to  go beyond the Planckian regime---in case such a thing  were possible. The history of physics is plagued, however, with examples where it is the unification of old concepts which led to the extension of  its limits of applicability. Thus, embracing this perception one  naively can write Eq.(\ref{cutoff}) as:
 $(-iqc^2/\hbar G^{1/2})_{\rm cutt \, off}  \sim l^{-1}_{pl}.$  Meaning that there is somewhere a feature of the space-time whose description is given by complex numbers and discrete structures, which might just control the divergences of the theory so that it can be extended its applicability a little bit further.  Likewise the other elastic constants of the metric should form a set of clues for the unification of the geometry with the quantum. And if one commits the terrible felony of  bringing out the measurement paradox: by making the bold assumption that the phenomenon of wavefunction collapse is a real physical process where  gravity  is involved; then, the original situation regarding the status of Einstein's gravity has been turned around, and  it  might, after all, provide us with important clues concerning the most basic principles of nature.   This might be an instance of the oft-repeated dictum of Niels Bohr\cite{Heisenberg}:
 
 ``The opposite of a correct statement is a false statement. But the opposite of a profound truth may well be another profound truth."  
 
 The mathematical formalism presented here seems to indicate that quantum mechanics is interconnected  with gravity in a subtle way, since under precise circumstances Einstein's general theory of relativity can be cast as a superconductivity theory of the four-dimensional spacetime; and furthermore,  geometrical boundary terms in the variational formulation of the theory, like the York-Hawking-Gibbons term $(2\chi)^{-1} \int 2 K,$ naturally leads to curious relationships closely akin to the Planck's radiation formula. 
 
    In the present article cold dark matter is linked to dark energy: the first was pictured as a quantum macroscopic phenomena, where the spacetime  acquires superconducting properties transporting vacuum energy---the analog of the electric charge---around galaxies and cluster of galaxies, while deforming the gravitomagnetic lines of force.  
    
     Gravitomagnetism: describing how local inertial frames are influenced and dragged by mass-energy currents relative to other masses, was predicted in the period of 1896-1916, and discussed even before the advent of  the complete formulation of the General theory of relativity\cite{WheelerCIU}.  This effect is so feebly in strength that it is not  easy to account for it by direct observation of natural celestial bodies in our solar system (i.e. Mercury going around the Sun,  Jupiter's fifth moon circling Jupiter).  However it has already been detected (the Lense-Thirring effect) by the LAGEOS satellites\cite{LAGEOS}.  Furthermore, on 24 April 2004, the Gravity Probe B spacecraft---equipped with superconducting gyroscopes spherical to one part in a million, and a star-tracking telescope---started collecting data about the gravitomagnestim of Earth for almost a year. The climax came on May 2011, the year of the centenary anniversary of the discovery of superconductivity, when the Gravity Probe B satellite experiment announced his final results in complete accordance with Einstein's 1916 theory of curved spacetime\cite{GP-B}. Thus, it is fair to assert that this approach to the dark matter conundrum---originated from the conviction that  gravity has a lot to say about why the quantum theory is the way it is---rest on this well founded aspect of physics.  

Take notice that essential features of the present theory appear---albeit in different form, in other approaches to the cold dark matter conundrum. For instance,  in the theory of modified gravity known as T$e$V$e$S, proposed by J. Bekenstein \cite{TeVeS},  it is resorted to a vector and a scalar field, as well as  conformal deformations of the metric; on the other hand, axionic cold dark matter is a theory about Pseudo-Nambu-Goldstone bosons. 

Let us point out, that standard (collisionless) cold dark matter seems to have problems on small scales: due to cuspy central density profile haloes that are in conflict with observation of dwarf galaxies\cite{Merrit}, due to a delicate issue of the mass growth rates of central black holes through the capture of cold dark matter particles\cite{Hernandez}. Self-interacting Bose-Einstein-condensates  (BEC) for cold dark matter  have been offered as an alternative, since they can produce galactic halos with constant density cores. In the case of axionic dark matter, it has been argued that the formation of quantum vortices by the superfluidity of the BEC cold dark matter halo is not expected,  since axions are effectively non interacting; vortices however will be created in strongly-coupled condensates\cite{Tanja}. This is a relevant fact since in this article some estimates for the formation of spacetime quantum vortices have been given. The proposed theory also predicts the expulsion of  gravitomagnetic fields in analogy with the Meissner effect.  

The corresponding  microscopic theory of superconductivity of the spacetime has  not yet been provided. The result stands only at a phenomenological level, in terms of the Gizburg-Landau equations of superconductivity, which however seem to provide a route worth exploring for the explanation of  a real physical phenomenon occurring in the outskirts of galaxies.  It is worth mentioning that the BCS theory was developed almost half a century after the discovery of superconductivity in metals, and less than a decade after the appearance of the famous paper by Gizburg and Landau. Then, in 1986, by adding barium to crystals of lanthanum-copper-oxide, Bednorz and Muller\cite{Bednorz} discovered high-temperature superconductivity (HTS) in ceramic materials. The mechanism behind HTS, however, is still obscure.  Several theoretical schemes to explain HTS has been proposed, including  BCS-like theories (excitonic, plasmonic, magnetic, kinetic), the bipolaron theory, and the  resonating valence bond theory (RVB). For instance, in the RVB theory proposed by 
P. W. Anderson\cite{Anderson}, the fundamental entities for making the flow behave as if the electron had broken apart into separate particles,  one containing its charge but having no spin---the holon, and one carrying its spin but having no charge---the spinon. Thus, further thought will be required as there are serious choices to make to propose a reasonable microscopic theory for spacetime superconductivity. 

  Here we showed that, quantum mechanically, the fabric of the spacetime can act as if it were a type II superconductor characterized by the Gizburg-Landau parameter
$\kappa=1.5.$ This result can be regarded as a direct manifestation of the wave-particle duality of gravitation.      
 
``The bucket water experiment illustrating Mach's principle here on Earth has gone wild in the heavens, the `swimming pool'  where 
 galaxies float apparently  is not filled with an ordinary classical substance, but with a spacetime superfluid, charged with vacuum energy, to trick us all.''
 \begin{acknowledgments}
The author would like to express his gratitude to  M. Ramos-D\'iaz  for support, to  J. Guven for earlier discussions on the Lichnerowicz equation, and to the UC MEXUS-CONACyT organization for a postdoctoral fellowship under grant $040013.$ \\
\end{acknowledgments} 
   
 \end{document}